\documentclass[12pt,preprint]{aastex}

\newcommand{\mkms}{{\rm \; km\;s^{-1}}}
\newcommand{\kms}{km~s$^{-1}$}
\newcommand{\dnzmgg}{$dN/dz|_{\rm GRB}$}
\newcommand{\dnzmgq}{$dN/dz|_{\rm QSO}$}
\newcommand{\mgii}{\ion{Mg}{2}}

\shorttitle{Weak \ion{Mg}{2} in GRB sightlines}
\shortauthors{Tejos, Lopez, Prochaska et al.}

\begin{document}

\title{Casting light on the 'anomalous' statistics of \mgii\ absorbers
  toward Gamma-Ray Burst afterglows: the incidence of weak systems}


\author{Nicolas Tejos\altaffilmark{1}, 
        Sebastian Lopez\altaffilmark{1}, 
        J. Xavier Prochaska\altaffilmark{2},
	Joshua S. Bloom\altaffilmark{3},
        Hsiao-Wen Chen\altaffilmark{4}, 
	Miroslava Dessauges-Zavadsky\altaffilmark{5} and
	Maria J. Maureira\altaffilmark{1}
}

\altaffiltext{1}{Departamento de Astronom\'{\i}a, Universidad de
Chile, Casilla 36-D, Santiago, Chile; ntejos@das.uchile.cl}

\altaffiltext{2}{Department of Astronomy, UCO/Lick
Observatory, University of California, 1156 High Street, Santa Cruz,
CA 95064}

\altaffiltext{3}{Department of Astronomy, 601 Campbell Hall,
University of California, Berkeley, CA 94720-3411}

\altaffiltext{4}{Department of Astronomy, University of Chicago, 5640
S. Ellis Ave., Chicago, IL 60637}

\altaffiltext{5}{Geneva Observatory, University of Geneva, 51, Ch. des
Maillettes, 1290 Sauverny, Switzerland}



\begin{abstract}

We revisit echelle spectra (spectral resolution $R \approx 40\,000$)
of $8$ Gamma-Ray Burst afterglows to obtain the incidence ($dN/dz$) of
weak intervening \mgii\ systems at a mean redshift of $\langle z
\rangle = 1.5$. We show that $dN/dz$ of systems having restframe
equivalent widths $0.07$ \AA~ $ \le W_r^{{\rm MgII}} < 1$ \AA\ toward
GRBs is statistically consistent with the incidence toward
QSOs. Our result is in contrast to the results for \mgii\ systems
having $W_r \ge 1$ \AA, where $dN/dz$ toward GRBs has been found to be
larger than toward QSOs by a factor of $\approx 4$. We confirm
the overdensity albeit at a factor of $\approx 3$ only. This suggests
that any explanation for the GRB/QSO discrepancy, be it intrinsic to
the absorbers or a selection effect, should be inherent only to the
galaxies that host strong absorbers in the line-of-sight to
GRBs. We argue that, of all scenarios that have been proposed, lensing
amplification is the one that could explain the strong \ion{Mg}{2}
enhancement while allowing for no significant enhancement in the weak
absorbers.

\end{abstract}



\section{Introduction}

The recent refinements in rapid-response spectroscopy of high-redshift
Gamma-Ray Burst (GRB) optical afterglows have opened a new era in the
study of the intergalactic medium
\citep[IGM,][]{vel+04,fiore05,chen05}. In fact, using GRB afterglows
instead of quasi-stellar objects (QSO) as background sources
represents a superb complement to the absorption line technique
in terms of redshift coverage \citep[the highest redshift objects
detected are GRBs, e.g.,][]{olivares09,tanvir09}, ease of absorption
system identification (no emission lines in afterglow spectra), and
new insights into the interstellar medium of the host galaxies 
\citep{savaglio06,pcd+07}, not to mention the novel access to the
absorbers via deep imaging that the rapid fade-out of the afterglow
permits \citep[e.g.,][]{chen09,pcp+09}.

The first systematic, spectroscopic study of intervening systems
toward GRB afterglows delivered the first surprise. Using spectra
sensitive to restframe equivalent width (EW) $W_r^{2796}\ge 1$
\AA\ \mgii\ systems at a mean redshift of $\langle z \rangle=
1.1$, \citet[hereafter P06]{prochter06} identified $14$ such strong
systems in a sample of 14 afterglow spectra at velocities $\beta c >
3\,000 \mkms$ from the GRB redshift. The redshift-path covered yielded
almost $1$ strong \mgii\ system per unit redshift, a roughly $4$ times
higher incidence than toward QSO lines-of-sight at greater than
$99.9\%$ confidence. Since the intervening absorption systems are
thought to be physically independent of the background source,
this result has called for a serious revision of our understanding of
absorption line surveys.

Four main astrophysical effects have been proposed to explain the
observed discrepancy \citep[see P06;][]{porciani07,
cucchiara08,sss09}: strong \mgii\ gas might be intrinsic to the
GRB environment or host galaxy system; dust within strong
\mgii\ absorbers might obscure faint QSOs that never get detected; 
GRBs might be gravitationally lensed (and amplified) by the absorbers. 
A fourth scenario, namely that small absorber sizes might
make the distinct beam sizes of GRBs and QSOs affect the statistics
differentially \citep{frank07}, has proven to be unviable
\citep{pontzen07,thone08,aoki08}. However, as argued in P06
and \citet{porciani07}, none of these effects alone is likely to
explain the QSO/GRB discrepant \mgii\ statistics. More recent studies
have shown that the \ion{C}{4} statistics of QSOs and afterglows are
consistent with each other \citep{tejos07,sudilovsky07}, although
those surveys probed a much higher redshift and also probably
different galactic environments.

In this paper we use echelle spectra of GRB afterglows, sensitive to
$W_r^{2803}\ge 0.07$ \AA, to explore the {\it weak} \mgii\
systems. The QSO \mgii\ EW distribution shows a clear turnover around
$W_r \sim 0.3$ \AA, hinting at different populations \citep[e.g.,][
hereafter N07]{churchill99,nestor05,milutinovic06,narayanan07}.  Here
we show for the first time that, contrary to the strong systems, the
{\it weak} ($W_r < 0.3$ \AA) and the {\it moderately strong} ($0.3 \le
W_r < 1$ \AA) \ion{Mg}{2} statistics conform to those derived from QSO
surveys. In view of these new results we discuss possible explanations
for the P06 result.

\section{Data and Search Algorithm}\label{data}
The GRB afterglow sample comprises 8 echelle optical spectra ($R
\equiv \lambda/\delta\lambda\approx 40\,000$ and $S/N > 5$ pix$^{-1}$)
taken with the Keck/HIRES \citep{hires}, Magellan/MIKE \citep{mike}
and VLT/UVES \citep{uves} spectrographs. This dataset comprises all
current GRB echelle spectra available to our
group. Table~\ref{GRBsample} lists the targets that we have used,
along with references. Five of these spectra were used in the P06
survey (GRB021004, GRB050730, GRB050820, GRB051111, and GRB060418) and
three are new (GRB050922C, GRB060607, and GRB080810). Note also that
our survey extends beyond $z = 2$, while the P06 survey was restricted
to $z \le 2$.  Data reduction was conducted in the same fashion as
described in \citet{tejos07}.

To identify \ion{Mg}{2} systems in our sample we proceeded in two
steps. We first performed a blind and automatic search for absorption
lines using the "aperture method" \citep{wolfe86, churchill08}. This
yielded a list of lines detected at the $2.5\sigma$ confidence
level. \mgii\ doublet candidates were searched for in this list by
imposing a $5\sigma$ confidence limit on the stronger doublet $\lambda
2796$ line, but no constraint on the doublet ratio (DR) in order not
to exclude blended lines.

The second step was to calculate EW values. To this end, we used
direct pixel integration and, to conform to analysis techniques in QSO
surveys (e.g., N07), complex systems were considered as a single one
if the velocity span $\Delta v<500$ \kms. A careful inspection by eye
allowed us to exclude spurious systems and obvious blends, and the
final sample was built by imposing the criterion $1<{\rm DR}<2$. This
last condition did not exclude any possible system.

\begin{figure}[h]
\begin{center}
\includegraphics[angle=0,width=15cm]{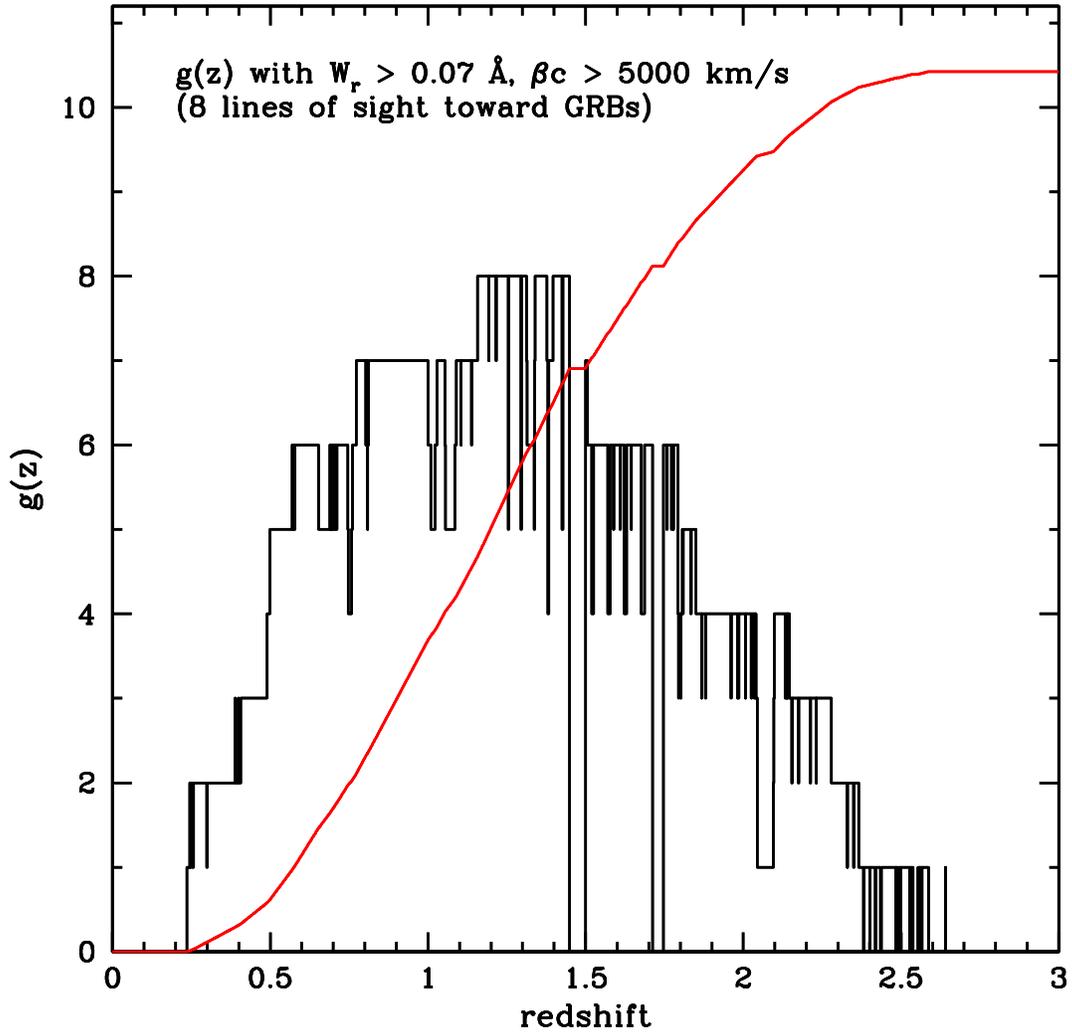}

\caption{Number of lines of sight in the \ion{Mg}{2}~ survey and
cumulative redshift path as a function of redshift for $W_{min}=0.07$
\AA.}  \label{gz_MgII} 

\end{center} 

\end{figure}

To test the sensitivity of our search algorithm we ran it over a
sample of synthetic spectra of signal-to-noise ratio $S/N=5$ and
containing \ion{Mg}{2} systems having a variety of column densities
and Doppler parameters. This $S/N$ ratio or better is representative
of $\gtrsim 90 \%$ of the redshift path, $\Delta z$. The efficiency
was inferred by counting how many doublets were recovered over the
total. The result of this analysis was that our detection method
recovers $100$\%\ of the lines having $W_r^{2803} \ge 0.07$ \AA\ at
this $S/N$ level. Thus, for our survey we take $W_{min} = 0.07$ \AA\
in {\it both} components of the doublet. For this limit, the total
redshift path is $\Delta z = 10.42$. The redshift-path density,
$g(z)$, is shown in Figure \ref{gz_MgII}.

\section{Sample Definitions}\label{sample}

We define following statistical samples:
\paragraph{Full Sample (FS)}
All \ion{Mg}{2} systems between the redshifted Ly$\alpha$ and
\ion{Mg}{2} associated with the GRB, with $W_r^{2803} \ge 0.07$~\AA\
and detected at the $5\sigma$ and $2.5\sigma$ confidence level in
$\lambda2796$ and $\lambda2803$, respectively.  The Full Sample is
composed of 23 \ion{Mg}{2} systems (listed in Table \ref{FULL}). Note that we 
did not find any system with $W_r^{2803} > 0.07$ \AA\ in
the GRB081010 spectrum and that \mgii-free sightlines are
expected from QSO surveys.

\begin{figure}[h]
\begin{center}
\includegraphics[angle=0,width=15cm]{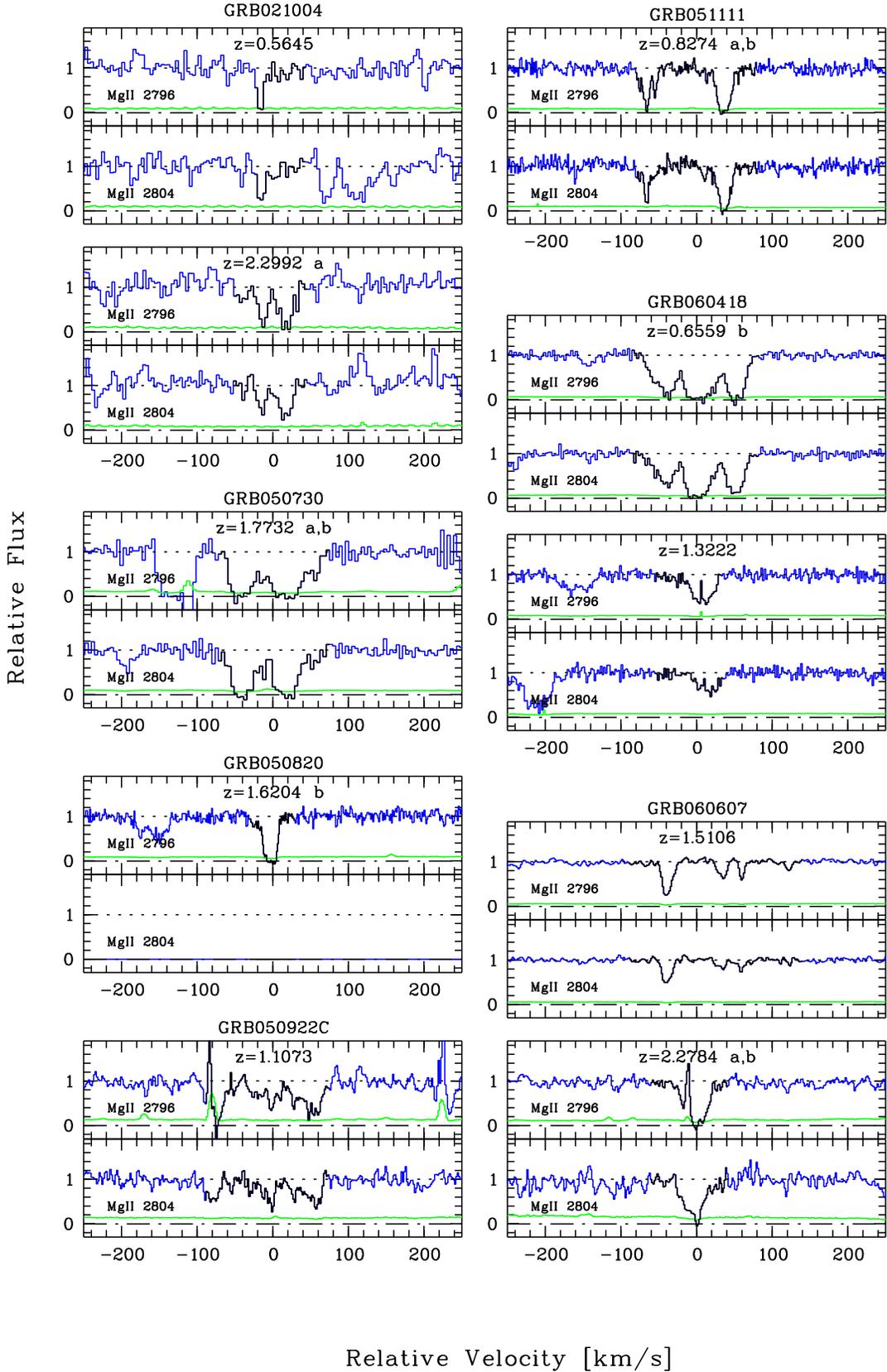}

\caption{Velocity profiles of intervening \ion{Mg}{2} systems with
  $W_r^{2796} < 1.0$ \AA. The labels to the right of the redshifts
  indicate a \ion{Mg}{1}~ ('a') or \ion{Fe}{2}~ ('b') detections.}
  \label{systems} \end{center} \end{figure}

\paragraph{Intervening Sample (IS)}
All systems in the FS but excluding those ones within $5\,000$ \kms\
of $z_{GRB}$ (labeled as 'Local' in Table \ref{FULL}). This
sample is composed of 19 systems, having a median redshift of $\langle
z \rangle = 1.4$.  Figure \ref{systems} shows the velocity profiles of
those systems with $W_r^{2796} < 1$~\AA.

\paragraph{Strong Intervening Sample (SIS)}
All systems in the IS having $W_r^{2796} \ge 0.3$ \AA. This is the
same cutoff used in QSO absorption line surveys
\citep[e.g.,][N07]{nestor05}. This sample is composed of 14 systems
(labeled as 'S' in Table \ref{FULL}) and is complete at the $99$\%
level along a redshift path of $\Delta z = 10.86$. Systems with
$W_r^{2796} \ge 1.0$ \AA\ are labeled with a 'V' (very strong) in the
table. This latter limit is the same used by P06.

\paragraph{Weak Intervening Sample (WIS)}
All systems in the IS having $W_r^{2796} < 0.3$ \AA.  This sample is
composed of 5 systems (labeled as 'W' in Table \ref{FULL}) and is
complete at the $96$\% level over $\Delta z = 10.42$. The QSO
absorption line survey we compare with was that one by N07. However,
these authors were able to use the more sensitive limit
$W_{min}^{2796}= 0.02$ \AA. Consequently, for the sake of comparison
between the GRB and QSO data, we recomputed \dnzmgq\ using a
sub-sample drawn from their line list.

\section{Results}\label{results} 

\begin{figure}[h]
\begin{center}
\plotone{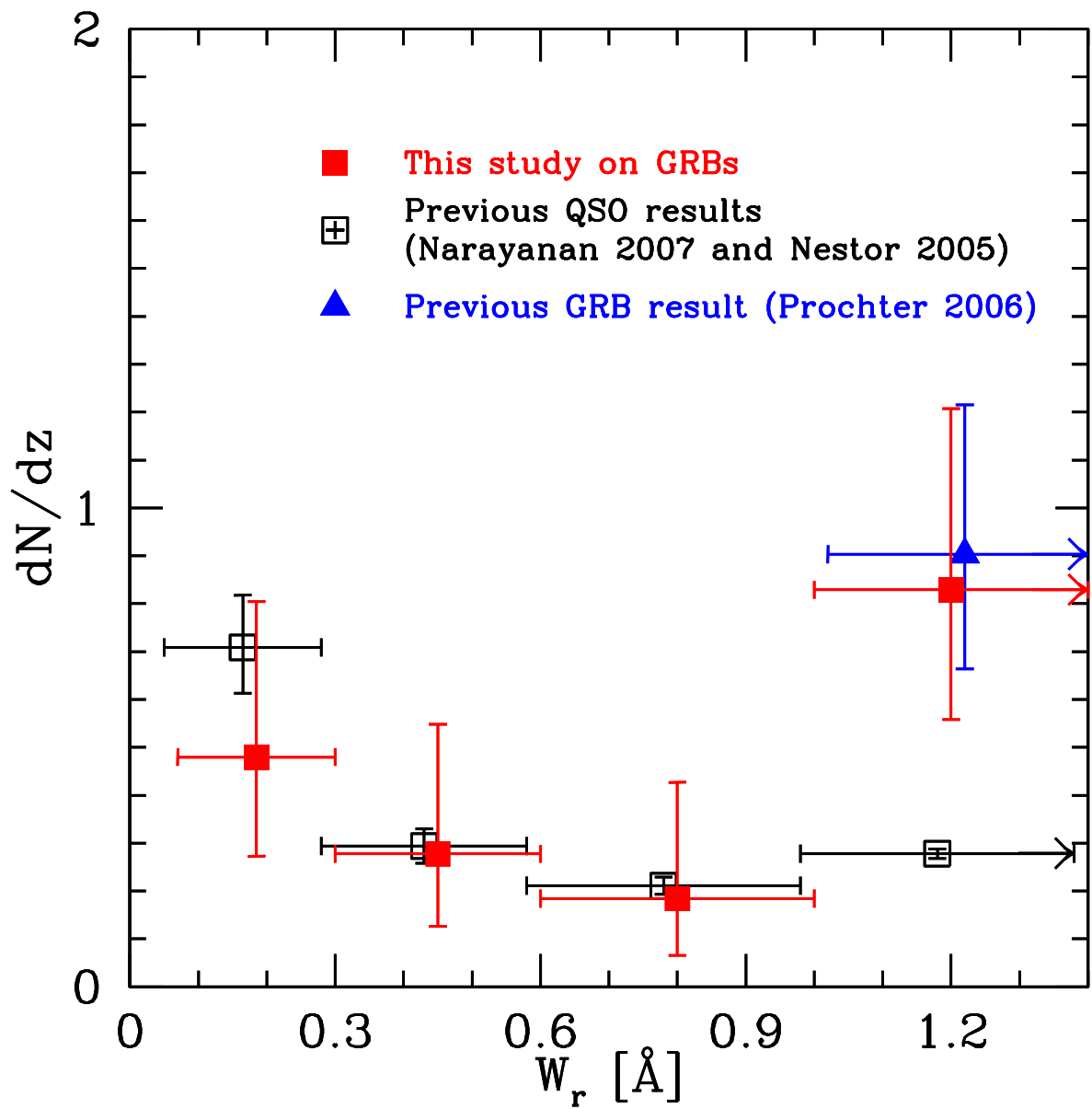}
\caption{Redshift number density of \ion{Mg}{2} absorption systems
  toward GRB afterglows (filled squares; see the numbers in Table
  \ref{resultMgII}). Empty squares (slightly offset in $x$ for the
  sake of clarity) depicts the QSO results from N07 ($W_r^{2796} <
  0.3$ \AA) and \citet{nestor05} [$W_r^{2796} \ge 0.3$ \AA]. The
  triangle indicates the P06 result for GRBs. Note that both
  surveys have 5 spectra in common and are therefore not completely
  independent. Also note that the high EW bin corresponds to
  $W_r^{2796} \ge 1$ \AA.}
  \label{fig_MgII_results_Wdist} \end{center} \end{figure}

Table \ref{resultMgII} shows \dnzmgg\ \citep[calculated in the same
fashion as in][]{tejos07} for 4 EW bins, along with \dnzmgq\ in the
same bins.  These numbers are plotted in Figure
\ref{fig_MgII_results_Wdist}.  At this point it is important to
emphasize that, as in \citet{tejos07}, our error estimation for
the Poisson statistics is based on the tables given by
\citet{gehrels86} for small numbers. These errors are larger than the
usual approximation, $\sigma_{N} = \sqrt{N}$.

From Figure \ref{fig_MgII_results_Wdist} it is clear that our results
for GRB sightlines match those ones for QSOs for equivalent widths
$W_r^{2796} < 1$ \AA, while for those with $W_r^{2796} \ge 1$ \AA~ we
recover a similar overabundance as found by P06 which included
low-resolution data.

In the WIS our result for GRB sightlines, \dnzmgg$(\langle z \rangle =
1.4 ) = 0.48^{+0.32}_{-0.21}$, is consistent with \dnzmgq$(\langle z
\rangle = 1.2 ) = 0.71 ^{+0.11}_{-0.10}$ that we infer from the data
presented by N07 (55 systems at $0.4<z<2.4$ having $W_r^{2803} \ge
0.07$ \AA~ and $W_r^{2796} < 0.3$ \AA\ in a total redshift path
$\Delta z = 77.6$\footnote{The slightly different redshift coverage
  between the N07 data and ours makes no significant difference in
  this comparison.}).  We find that our central value is actually
  $\approx 70\%$ of the incidence estimated for QSO sightlines, but
  this difference is not significant. Therefore, we consider an
overabundance of weak \ion{Mg}{2} systems in GRB sightlines compared
with that from QSOs to be very unlikely.

On the other hand, in the SIS we recover the result obtained by P06
for \mgii\ systems with $W_r^{2796} \ge 1$ \AA, although we find an
overabundance of a factor of 3 only, instead of 4, when comparing with
the QSO results by \citet{nestor05}.  Because our redshift path is
only two thirds that of P06, the significance of the result is reduced
from $99.9\%$ to $\approx 95.5\%$. Nonetheless, the fact that we have
added new lines-of-sight argues that the GRB/QSO discrepancy is real,
and possibly not due to statistical uncertainties nor a posteriori
subtleties. However, note that both surveys have 5 spectra in common
and are therefore not completely independent.

\begin{figure}[h]
\begin{center}
\plotone{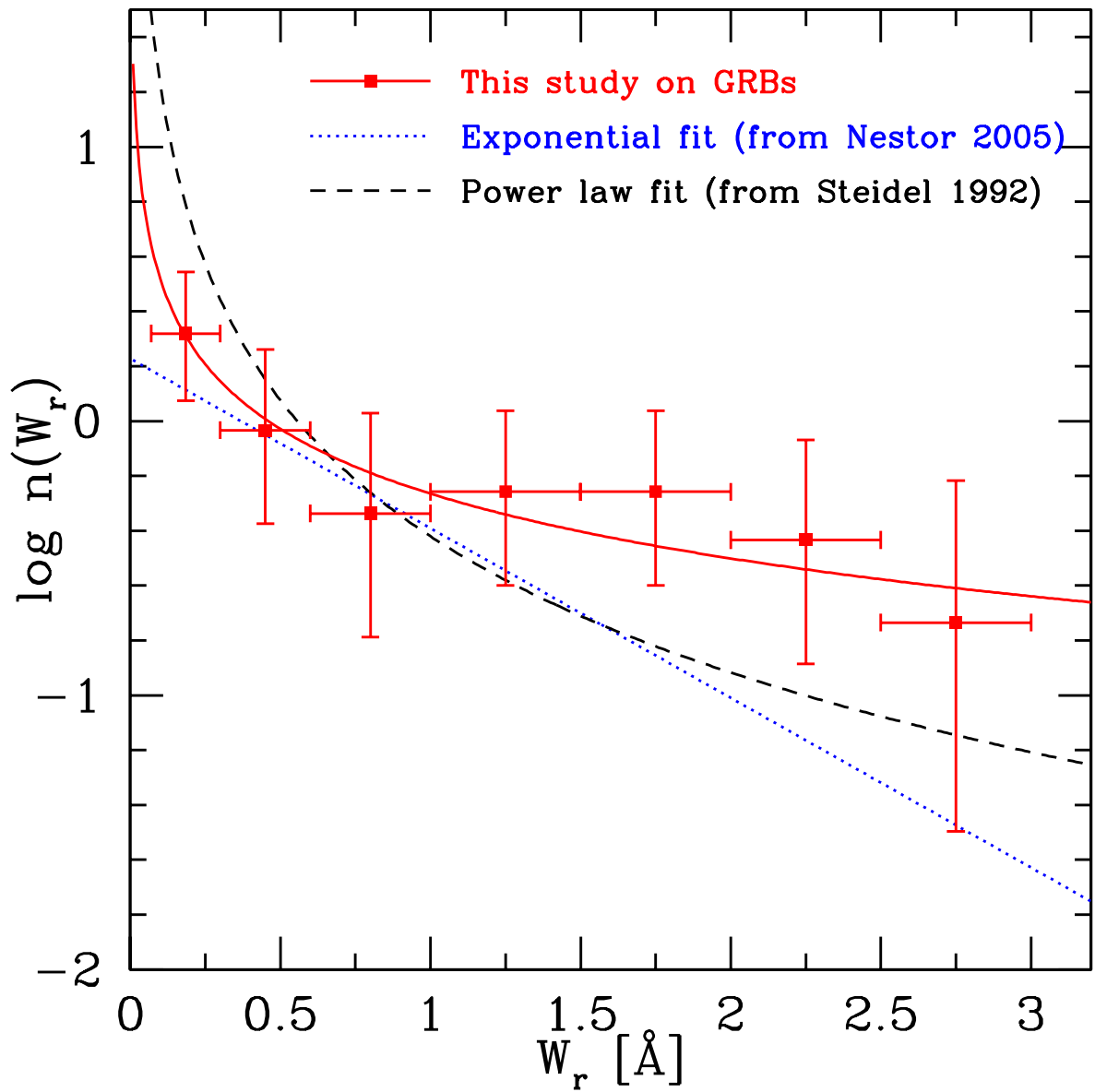}
\caption{Equivalent width distribution of \ion{Mg}{2} absorption
systems toward GRB afterglows (filled squares). The solid line
corresponds to the best-fit power-law to the data points. The dotted
and dashed lines correspond to the expected distributions from QSO
sightlines: an exponential fit for systems at $W_r^{2796} \ge 0.3$
\AA~ \citep[dotted line; from][]{nestor05} and a power-law fit for
systems at $W_r^{2796} < 0.3$ \AA~ \citep[dashed line;
from][]{steidel92}.}
\label{wdist}
\end{center}
\end{figure}

Finally, let us emphasize that there seems to be a transition at
$W_r^{2796} \ge 1$ \AA, as we see no significant GRB/QSO differences
for intermediate EW values ($0.3$ \AA~ $ \le W_r^{2796} < 1$
\AA). This is more clearly seen in Figure \ref{wdist}, which shows the
EW distribution in our GRB sample compared with previous
parameterizations obtained from QSO samples
\citep{nestor05,steidel92}.

\section{Discussion and Implications}\label{discuss}

The fact that we do not find any discrepancy between the statistics of
QSO and GRB weak \mgii\ systems opens the question as of why there is
an overabundance of systems only for $W_r^{2796} > 1$ \AA~ systems in
front of GRBs. Although the extant sample of afterglow spectra is
still small, our result suggests that any explanation for the GRB/QSO
discrepancy, be it intrinsic to the absorbers or a selection effect,
should be inherent only to the galaxies that host strong absorbers in
the line of sight to GRBs. In the following we discuss how the
different models proposed to explain the P06 result may or may not be
reinforced in light of our new results on weak systems.

\paragraph{Absorbers Intrinsic to the GRBs}
The present high-resolution spectra seem to rule out an intrinsic
origin of the \mgii\ systems for two reasons. First, the line
profiles, as seen at high spectral resolution, show no indication of
broad and shallow absorption troughs, characteristic of BAL
QSOs\footnote{However, we note that very shallow systems would not, 
in most cases, be detected in our GRB spectral 
sample.}. Secondly, if some of the \ion{Mg}{2} systems were intrinsic
to the GRB, we would expect an overabundance also for the $W_r < 1$
\AA~ \ion{Mg}{2} systems, which we do not observe \citep[indeed, an
overabundance of strong \ion{C}{4} would be expected too, and that is
also not observed;][]{tejos07}.


\paragraph{GRB and QSO Beam Sizes}
The geometrical model proposed by \citet{frank07} (based on different
GRB and QSO beam sizes, both comparable to the \ion{Mg}{2} absorber
characteristic sizes) has been tested and ruled out by subsequent
observational analysis \citep{pontzen07}.  Furthermore, initial claims
of line-strength variability from \citet{hao07} in a single sightline
(GRB060206) have been refuted \citep{thone08,aoki08}. Nevertheless, we
will consider this model in light of our new observations.

A consequence of the geometrical model \citep[see][]{porciani07} is
that a fraction of weak systems in QSO spectra should have $DR \approx
1$. From the N07 sample we find this fraction to be $\approx 5\%$. Due
to the smaller GRB beam sizes, the same fraction in GRB spectra is
expected to be lower than this value. In contrast, we find that 2 out
of 5 systems with $W_r^{2796} < 0.3$ \AA\ show $DR\approx1$ (note that
taking larger EW values would include saturated lines). Thus, this
number, though not significant, does not support the geometrical
model.

In addition, the model also predicts an underabundance of weak
systems. This is suggested by our data for $W_r^{2796} < 0.3$ \AA\
systems, but the $dN/dz$ values are consistent at the $1\sigma$
confidence level.

\paragraph{Dust}
As discussed in P06 and \citet{porciani07}, the apparent high
incidence of strong \mgii\ absorbers toward GRBs might be explained by
an underestimated incidence of strong \mgii\ systems toward QSOs, as a
consequence of sources that get lost due to dust obscuration.
Although there is mounting counterevidence for a dust bias in QSO
surveys \citep{ellison01,ellison09,menard08}, from the point of view
of the GRBs data alone our result on {\it weak} absorbers, a priori
does not rule out the dust-obscuration scenario, at least
qualitatively. This is so because dust is supposed not to have a
considerable obscuring effect when $W_r<1$ \AA\ \citep{menard08}.

On the other hand, a scenario where dust reduces the incidence of
strong systems only in QSO sightlines is puzzling.  In this scenario,
the GRBs provide the unbiased (i.e., 'real') EW distribution but the
observed EW distribution for GRBs is atypical (see Figure \ref{wdist})
when compared against any other line surveyed along QSO or GRB
sightlines \citep[e.g., \ion{C}{4}, Ly$\alpha$;][]{paschos08}. It
seems that there is a transition at $W_r^{2796} \ge 1$ \AA~ where
the EW distribution does not decrease as it would be expected.
Therefore, in view of our new results, we conclude that dust is
unlikely to explain the differences between \ion{Mg}{2} toward QSO and
GRB sightlines.

\paragraph{Gravitational Lensing}
Source amplification due to strong gravitational lensing may bias the
GRB spectral samples toward targets that contain more intervening
absorbers, if these occur in the lensing galaxies
\citep[P06;][]{porciani07}. Our spectral sample does not offer a
direct means to infer what kind of \mgii\ systems may be associated to
galaxy configurations being more or less lensing-efficient. Obviously,
further deep late-time imaging observations of GRB fields
\citep[e.g.,][]{chen09} must be carried out in order to identify the
absorbing galaxies and possibly look for
impact-parameter/line-strength correlations.

Nevertheless, if we speculate that the strong absorber overdensity is
purely explained by a selection effect due to lensing magnification,
our results can help us estimate the fraction $f_l$ of magnified GRBs
that otherwise would not have been spectroscopically observed. To
estimate $f_l$, let us consider a \mgii~survey composed by $M$ QSO
sightlines. Then, the expected number of absorption systems will be:

$$ N_{QSO} = \frac{dN}{dz}_{QSO} \langle \Delta z \rangle M \ {\rm,}$$

\noindent where $\langle \Delta z \rangle$ is the average redshift
path per sightline and $dN/dz|_{QSO}$ is the expected incidence of
systems (assumed unbiased; a quantitative detail of a possible lensing
bias in QSO surveys is beyond the scope of this paper).

Let us now consider an equivalent GRB survey with $M$ sightlines. If
the observed number of absorption systems, $N_{GRB}$, is a factor of
$e$ greater than $N_{QSO}$, then the excess of systems will be $N_{e}
= N_{QSO} (e - 1)$. Let $L$ be the total number of lensed sightlines
in that GRB survey. The fraction of magnified GRBs is then:

$$f_l \equiv \frac{L}{M} \ {\rm.}$$

\noindent If we assume that the excess of systems is just due to
lensing (either macro or microlensing\footnote{The optical depth for
microlensing increases at low impact parameters from galaxies (the
surface density of stars and MACHOs is greater in the center than in
the outskirts of galaxies) therefore it should also contribute to the
excess of strong systems \citep[see also][]{porciani07}.}), then any
extra system corresponds to a lensed sightline:

$$L = N_{e} \ {\rm,}$$

\noindent and therefore:

$$f_l = (e-1) \frac{dN}{dz}_{QSO} \langle \Delta z \rangle \
{\rm.}$$

Thus, in order to reproduce the factor of $\approx 3$ enhancement that
is observed at this EW level, we estimate that $f_l$ must be of the
order of $\approx 60\%$ ($e \approx 3$, $dN/dz|_{QSO} \approx 0.3$ and
$\langle \Delta z \rangle \approx 1$). Such a fraction would add twice
as many strong systems as encountered if there were no
lensing\footnote{Note that this argument becomes unrealistic for a
factor of $\approx 4$ enhancement, for which $f_l$ would approach
$\approx 100\%$.}. Similarly, an enhancement factor of $\approx 2$
(still consistent with our result at the 1$\sigma$ c.l.) would require
$f_l \approx 30\%$. Since more realistically $L \le N_e$, this
estimate of $f_l$ should be taken as an upper limit. Note that we do
not provide here a quantitative assessment of the lensing
magnification but instead we assume that it is large enough to provide
$f_l > 0$.  In fact, in the above situation our results would imply
that the lensing agents contribute only systems with $W_r \ge 1$ \AA\
(where $e>1$; note that this could be easily explained if weak
absorbers were indeed more external to galaxies, as proposed by N07
among others). In summary, we believe that lensing by the galaxies
hosting strong absorbers provides a viable explanation to the QSO/GRB
discrepancy \citep[see also][]{vergani09}.

%

  




 
A test of the lensing hypothesis could be made with very rapid and
deep spectroscopy of 'dark' bursts \citep[e.g.,][]{perley09}, for
which \dnzmgg\ should show no enhancement. In addition, as mentioned
above, another test of this bias is that there should be more massive
(and more luminous) intervening galaxies at low impact parameters in
sightlines where the \mgii\ EW is larger.

\section{Summary}

We have used echelle spectra of 8 GRB afterglows, three of them new,
to show that the incidence of weak \mgii\ systems ($0.07$ \AA~$ \le
W_r^{{\rm MgII}} <1$ \AA) is the same as toward QSO lines of
sight. There seems to be a transition at $W_r \approx 1$ \AA, above
which \dnzmgg\ rises significatively to a factor of a few with respect
to \dnzmgq, as found by P06.  In view of the present results on weak
absorbers, we suggest that the GRB/QSO discrepancy should arise in the
galaxies that host the strong absorbers. Effects associated to the GRB
phenomenon like ejected absorbers or different beam-sizes are not
supported by the data presented here nor a selection effect due to
dusty absorbers.  Instead, of all effects proposed in the literature,
a bias toward sources amplified by lensing seems to be in best
agreement with our findings.

This paper includes data obtained through the Gamma-ray Bursts
Afterglows as Probes (GRAASP) Collaboration (http://www.graasp.org) from
the following observatories: the W. M. Keck Observatory, which is a
joint facility of the University of California, CIT, and NASA, and
the 6.5 m Magellan Telescopes located at Las Campanas Observatory,
Chile. This paper also includes data based on observations made with
ESO Telescopes at the Paranal Observatories under programs
070.A-0599(B), 075.A-0603(B) and 077.D-0661(A). SL and NT are partly
supported by the Chilean {\sl Centro de Astrof\'\i sica} FONDAP
No. 15010003, and by FONDECYT grant N$^{\rm o}1060823$.  JXP is
partially supported by NASA/Swift grant NNG05GF55G and an NSF CAREER
grant (AST-0548180).

\newpage

\begin{deluxetable}{lllllcc}
\tablecolumns{7} 

\tablecaption{GRB Spectral Sample.}

\tablewidth{0pt}

\tablehead{
  \colhead{Source}&
  \colhead{RA (J2000)} &
  \colhead{Dec (J2000)} &
  \colhead{$z_{GRB}$}&
  \colhead{$\Delta z ^{a}$} &
  \colhead{Instrument}&
  \colhead{Reference}
  
}
\startdata

GRB021004 & 00 26 54.68 & +18 55 41.6 &2.335 & 1.55  &  UVES    & 1\\      
GRB050730 & 14 08 17.14 & $-$03 46 17.8 &3.969 & 1.14  &  MIKE    &2,3 \\ 
GRB050820 & 22 29 38.11 & +19 33 37.1 &2.615 & 1.10  &  HIRES   &3 \\
GRB050922C& 21 09 33.30 & $-$08 45 27.5 &2.199 & 1.56  &  UVES    & 4\\
GRB051111 & 23 12 33.36 & +18 22 29.5 &1.549 & 0.93  &  HIRES   & 3\\
GRB060418 & 15 45 42.40 & $-$03 38 22.8 &1.490 & 1.18  &  MIKE    &  3\\
GRB060607 & 21 58 50.40 & $-$22 29 46.7 &3.082 & 1.61  &  UVES    & 5 \\ 
GRB080810 & 23 47 10.40 & +00 19 11.0 &3.35  & 1.40  &  HIRES   & 6,7  
\enddata

\tablecomments{$^{a}$ Individual redshift path for our \mgii \ Intervening
Sample (see definition in \S \ref{sample}).}

\tablerefs{(1) \citet{fiore05}; (2) \citet{chen05}; (3)
\citet{prochaska07}; (4) \citet{piranomonte07}; (5) \citet{ledoux06};
(6) \citet{prochaska08}; (7) \citet{page08}.} \label{GRBsample}

\end{deluxetable}

\tabletypesize{\scriptsize}

\begin{deluxetable}{lccclccc}
\tablecolumns{8}
\tablecaption{Full Sample (FS) of \ion{Mg}{2} \ in GRB Spectra}
\tablewidth{0pt}
\tablehead{
  \colhead{GRB} &
  \colhead{z$_{GRB}$} &
  \colhead{z$_{start} $} &
  \colhead{z$_{end}^{\beta c=5000 km~s^{-1}} $} &
  \colhead{z$_{abs}^{MgII}$} &
  \colhead{W$_r^{pixel}$(2796)} &
  \colhead{DR} &
  \colhead{Comment$^{a}$}\\
  \colhead{}&
  \colhead{}&
  \colhead{}&
  \colhead{}&
  \colhead{}&
  \colhead{W$_r^{pixel}$(2803)} &
  \colhead{} &
  \colhead{}
}
\startdata

021004 & 2.335	& 0.44795  &   2.27938 & 0.56446&  0.140 $\pm$ 0.013   &   1.085  $\pm$ 0.155&  W\\
       &	&	   &   &             &     0.129 $\pm$ 0.014   &       & \\
       &        &          &   & 1.38067&          1.633 $\pm$ 0.016   &   1.251  $\pm$ 0.023     &    VS\\
       &	&	   &   &             &     1.305 $\pm$ 0.020   &       &\\
       &        &          &   & 1.60274&          1.389 $\pm$ 0.026   &   1.227  $\pm$ 0.025     &   VS\\
       &	&	   &   &             &	   1.132 $\pm$ 0.010   &       &\\
       &        &          &   & 2.29920  &        0.335 $\pm$ 0.016   &   1.701  $\pm$ 0.139     &   S\\
       &	&	   &   &             &	   0.197 $\pm$ 0.013   &        &  \\
       &        &          &   & 2.32893 &         0.814 $\pm$ 0.078   &   1.040 $\pm$ 0.108     &   Local\\
       &	&	   &   &             &	   0.783 $\pm$ 0.032   &       &\\

050730 & 3.97	& 1.5781   & 3.88711&  1.77317 &   0.923 $\pm$ 0.019   &   1.165  $\pm$ 0.038&  S\\
       &	&	   &   &             &	   0.792 $\pm$ 0.020   &       &\\
       &        &          &   & 2.25378     &   1.125 $\pm$ 0.034   &   1.426  $\pm$ 0.087&  VS\\
       &	&	   &   &             &     0.789 $\pm$ 0.042   &       &\\

050820 &  2.6147& 0.5694   & 2.55441  & 0.69153&   2.988 $\pm$ 0.022   &   1.280  $\pm$ 0.017&   VS\\
       &	&	   &   &             &	   2.335 $\pm$ 0.025   &       &\\
       &        &          &   &  1.43000     &	      \nodata        &     \nodata    & VS\\
       &	&	   &   &             &	   1.262 $\pm$ 0.016   &      & \\
       &        &          &   &  1.62040    &     0.226 $\pm$ 0.007   &   \nodata &  W\\
       &	&	   &   &             &	 \nodata         &          &\\

050922C& 2.199  & 0.3889 & 2.14565& 1.10731       &0.532 $\pm$ 0.031   & 1.482  $\pm$ 0.119&     S\\
       &	&	   &   &                  &0.359 $\pm$ 0.020   &       &\\
       &        &          &   & 2.19950    &      1.062 $\pm$ 0.058   & 0.700 $\pm$ 0.041      & Local\\
       &	&	   &   &             &	   1.518 $\pm$ 0.035   &       &\\

051111 & 1.549  & 0.1067   & 1.50649  &0.82735&    0.369 $\pm$ 0.010   &  1.242  $\pm$ 0.060&    S\\
       &	&	   &   &             &     0.297 $\pm$ 0.012   &       &\\
       &        &          &   & 1.18910&          2.091 $\pm$ 0.011   &  1.201  $\pm$ 0.009&          VS\\
       &	&	   &   &             &	   1.741 $\pm$ 0.010   &       &\\
       &        &          &   &1.54913     &      2.343 $\pm$ 0.010   &  1.106 $\pm$ 0.007     & Local\\
       &	&	   &   &             &	   2.118 $\pm$ 0.010   &       &\\

060418 & 1.49   & 0.0811   & 1.44847  &0.60259  &  1.299 $\pm$ 0.015   &  1.054  $\pm$ 0.018&VS\\
       &	&	   &   &             &     1.233 $\pm$ 0.015   &       &\\
       &        &          &   & 0.65593   &       0.975 $\pm$ 0.010   &  1.237  $\pm$ 0.021&       S\\
       &	&	   &   &             &     0.788 $\pm$ 0.011   &       &\\
       &        &          &   & 1.10724    &      1.832 $\pm$ 0.020   &  1.222  $\pm$ 0.019&      VS\\
       &	&	   &   &             &     1.499 $\pm$ 0.017     &      &\\
       &        &          &   & 1.32221    &      0.214 $\pm$ 0.009   &  1.609  $\pm$ 0.139&      W\\
       &	&	   &   &             &     0.133 $\pm$ 0.010   &       &\\
       &        &          &   & 1.48964    &      1.968 $\pm$ 0.017   &  1.150$\pm$  0.015   &   Local\\
       &	&	   &   &             &     1.711 $\pm$ 0.017   &    &\\

060607 &  3.082 & 0.7723   & 3.01392  & 1.51057&   0.197 $\pm$ 0.010   &  1.698  $\pm$ 0.183&   W\\
       &	&	   &   &             &	   0.116 $\pm$ 0.011   &       &\\
       &        &          &   &1.80208  &         1.906 $\pm$ 0.011   &  1.228  $\pm$ 0.015&         VS\\
       &	&	   &   &             &     1.552 $\pm$ 0.016   &       &\\
       &        &          &   & 2.27840     &  0.293 $\pm$ 0.015&  1.024  $\pm$ 0.080&  W\\
       &	&	   &   &             &	   0.286 $\pm$ 0.017   &       & \\

080810 &  3.35  & 0.107    &  3.277          &  \nodata  & \nodata & \nodata & 
\enddata

\tablecomments{All Mg II absorption systems between the redshifted Ly$\alpha$ and Mg II
 associated with the GRB. We did not find any system having
 $W_r^{2803} > 0.07$ \AA\ in the GRB081010 spectrum.\\ $^{a}$ See notation
 defined in \S 3.}  \label{FULL} \end{deluxetable} 


\begin{deluxetable}{lllll|l}
\tablecolumns{6} 

\tablecaption{Incidence of Intervening \mgii \ Absorption Systems
toward GRBs} \tablewidth{0pt}

\tablehead{
  \colhead{W$_r$ [\AA]}&
  \colhead{N$_{abs}$} &
  \colhead{$\Delta z$} &
  \colhead{$\langle z \rangle$}&
  \colhead{$dN/dz|_{\rm GRB}$} &
  \colhead{$dN/dz|_{\rm QSO}^{MgII \ a}$} 
  
}
\startdata
0.07 $\leq W_r^{2803}$ and $W_r^{2796} <0.3$ & 5 & 10.42 & 1.46 & 0.48 $^{+0.32}_{-0.21}$ & 0.71 $^{+0.11}_{-0.10}$\\
0.3  $\leq W_r^{2796} <$ 0.6        & 3    & 10.80       & 1.41 & 0.28 $^{+0.27}_{-0.15}$ & 0.29 $\pm$ 0.04 \\
0.6  $\leq W_r^{2796} <$ 1.0 & 2    & 10.86              & 1.22 & 0.18 $^{+0.24}_{-0.12}$ & 0.21 $\pm$ 0.02 \\
1.0  $\leq W_r^{2796}$         & 9  & 10.86              & 1.34 & 0.83 $^{+0.38}_{-0.27}$ & 0.28 $\pm$ 0.01 \\

\enddata 

\tablecomments{$^{a}$ Values in this column where calculated from N07
($W_r<0.3$ \AA) and \citet{nestor05} ($W_r \ge 0.3$ \AA). Since no
redshift list is available in \citet{nestor05}, we calculated $dN/dz$
assuming similar redshift coverage and $\frac{dN}{dz}(W_r^a<W_r<W_r^b)
= \frac{dN}{dz}(W_r^a<W_r) - \frac{dN}{dz}(W_r^b<W_r)$.}
\label{resultMgII} \end{deluxetable}

\clearpage

\begin{thebibliography}{}
\bibitem[Aoki et al.(2008)]{aoki08} Aoki, K., et al.\ 2008,
arXiv:0808.4157

\bibitem[Bernstein et al.(2003)]{mike} Bernstein, R., Shectman, S.~A.,
Gunnels, S.~M., Mochnacki, S., \& Athey, A.~E.\ 2003, \procspie, 4841,
1694

\bibitem[Chen et al.(2005)]{chen05} Chen, H.-W., Prochaska, J.~X.,
Bloom, J.~S., \& Thompson, I.~B.\ 2005, \apjl, 634, L25

\bibitem[Chen et al.(2009)]{chen09} Chen, H.-W., et al.\ 2009, \apj,
691, 152

\bibitem[Churchill et al.(1999)]{churchill99} Churchill, C.~W., 
Rigby, J.~R., Charlton, J.~C., \& Vogt, S.~S.\ 1999, \apjs, 120, 51

\bibitem[Churchill(2008)]{churchill08} Churchill, C. 2008, QSO
Absorption Lines Studies: Ultraviolet and Optical Spectroscopy

\bibitem[Cucchiara et al.(2008)]{cucchiara08} Cucchiara, A., Jones,
T., Charlton, J.~C., Fox, D.~B., Einsig, D., \& Narayanan, A.\ 2008,
arXiv:0811.1382

\bibitem[Dekker et al.(2000)]{uves} Dekker, H., D'Odorico, S., Kaufer,
A., Delabre, B., \& Kotzlowski, H.\ 2000, \procspie, 4008, 534

\bibitem[Ellison et al.(2001)]{ellison01} Ellison, S.~L., Yan, L.,
Hook, I.~M., Pettini, M., Wall, J.~V., \& Shaver, P.\ 2001, \aap, 379,
393

\bibitem[Ellison \& Lopez(2009)]{ellison09} Ellison, S.~L., \& Lopez,
S.\ 2009, arXiv:0904.3330

\bibitem[Fiore et al.(2005)]{fiore05} Fiore, F., et al.\ 2005, \apj,
624, 853

\bibitem[Frank et al.(2007)]{frank07} Frank, S., Bentz, M.~C., Stanek,
K.~Z., Mathur, S., Dietrich, M., Peterson, B.~M., \& Atlee, D.~W.\
2007, \apss, 312, 325

\bibitem[Gehrels(1986)]{gehrels86} Gehrels, N.\ 1986, \apj, 303, 336

\bibitem[Hao et al.(2007)]{hao07} Hao, H., et al.\ 2007, \apjl, 659,
L99

\bibitem[Kann et al.(2006)]{kann06} Kann, D.~A., Klose, S., \& Zeh,
A.\ 2006, \apj, 641, 993

\bibitem[Ledoux et al.(2006)]{ledoux06} Ledoux, C., Vreeswijk, P.,
Smette, A., Jaunsen, A., \& Kaufer, A.\ 2006, GRB Coordinates Network,
5237, 1

\bibitem[M{\'e}nard et al.(2008)]{menard08} M{\'e}nard, B., Nestor,
D., Turnshek, D., Quider, A., Richards, G., Chelouche, D., \& Rao, S.\
2008, \mnras, 385, 1053

\bibitem[Milutinovi{\'c} et al.(2006)]{milutinovic06} Milutinovi{\'c},
N., Rigby, J.~R., Masiero, J.~R., Lynch, R.~S., Palma, C., \&
Charlton, J.~C.\ 2006, \apj, 641, 190

\bibitem[Narayanan et al.(2007)]{narayanan07} Narayanan, A., Misawa,
T., Charlton, J.~C., \& Kim, T.-S.\ 2007, \apj, 660, 1093 (N07)

\bibitem[Nestor et al.(2005)]{nestor05} Nestor, D.~B., Turnshek,
D.~A., \& Rao, S.~M.\ 2005, \apj, 628, 637

\bibitem[Olivares et al.(2009)]{olivares09} Olivares, F., Kruehler,
T., Greiner, J., \& Filgas, R.\ 2009, GRB Coordinates Network, 9215

\bibitem[Page et al.(2008)]{page08} Page, K.~L., et al.\ 2008, GRB
Coordinates Network, 8080, 1

\bibitem[Paschos et al.(2008)]{paschos08} Paschos, P., Jena, T.,
Tytler, D., Kirkman, D., \& Norman, M.~L.\ 2008, arXiv:0802.3730

\bibitem[Perley et al.(2009)]{perley09} Perley, D.~A., et al.\ 2009,
arXiv:0905.0001

\bibitem[Piranomonte et al.(2006)]{piranomonte07} Piranomonte, S.,
D'Elia, V., Ward, P., Fiore, F., \& Meurs, E.~J.~A.\ 2006, Nuovo
Cimento B Serie, 121, 1561

\bibitem[{{Pollack} {et~al.}(2009){Pollack}, {Chen}, {Prochaska}, \&
  {Bloom}}]{pcp+09}
{Pollack}, L.~K., {Chen}, H.~., {Prochaska}, J.~X., \& {Bloom}, J.~S. 2009,
  ArXiv e-prints

\bibitem[Pontzen et al.(2007)]{pontzen07} Pontzen, A.,
Hewett, P., Carswell, R., \& Wild, V.\ 2007, \mnras, 381, L99

\bibitem[Porciani et al.(2007)]{porciani07} Porciani, C., Viel, M., \&
Lilly, S.~J.\ 2007, \apj, 659, 218

\bibitem[Prochaska et al.(2007)]{prochaska07} Prochaska, J.~X., et
al.\ 2007, \apjs, 168, 231

\bibitem[{{Prochaska} {et~al.}(2007){Prochaska}, {Chen},
  {Dessauges-Zavadsky}, \& {Bloom}}]{pcd+07} {Prochaska}, J.~X.,
  {Chen}, H.-W., {Dessauges-Zavadsky}, M., \& {Bloom}, J.~S.  2007,
  \apj, 666, 267

\bibitem[Prochaska et al.(2008)]{prochaska08} Prochaska, J.~X.,
Perley, D., Howard, A., Chen, H.-W., Marcy, G., Fischer, D., \&
Wilburn, C.\ 2008, GRB Coordinates Network, 8083

\bibitem[Prochter et al.(2006)]{prochter06} Prochter, G.~E., et al.\
2006, \apjl, 648, L93 (P06)

\bibitem[{{Savaglio}(2006)}]{savaglio06} {Savaglio}, S. 2006, New
Journal of Physics, 8, 195

\bibitem[Steidel \& Sargent(1992)]{steidel92} Steidel, C.~C., \&
Sargent, W.~L.~W.\ 1992, \apjs, 80, 1


\bibitem[Sudilovsky et al.(2007)]{sudilovsky07} Sudilovsky, V.,
Savaglio, S., Vreeswijk, P., Ledoux, C., Smette, A., \& Greiner, J.\
2007, \apj, 669, 741

\bibitem[Sudilovsky et al.(2009)]{sss09} Sudilovsky, V., Smith, D., \&
Savaglio, S.\ 2009, arXiv:0904.3227

\bibitem[Tanvir et al.(2009)]{tanvir09} Tanvir N., et al.\ 2009, GRB
Coordinates Network, 9219

\bibitem[Tejos et al.(2007)]{tejos07} Tejos, N., Lopez, S., Prochaska,
J.~X., Chen, H.-W., \& Dessauges-Zavadsky, M.\ 2007, \apj, 671, 622

\bibitem[Th{\"o}ne et al.(2008)]{thone08} Th{\"o}ne,
C.~C., et al.\ 2008, \aap, 489, 37

\bibitem[Vergani et al.(2009)]{vergani09} Vergani, S.~D., Petitjean,
P., Ledoux, C., Vreeswijk, P., Smette, A., \& Meurs, E.~J.~A.\ 2009,
\aap, 503, 771

\bibitem[{{Vreeswijk} {et~al.}(2004){Vreeswijk}, {Ellison}, {Ledoux}, {Wijers},
  {Fynbo}, {M{\o}ller}, {Henden}, {Hjorth}, {Masi}, {Rol}, {Jensen}, {Tanvir},
  {Levan}, {Castro Cer{\'o}n}, {Gorosabel}, {Castro-Tirado}, {Fruchter},
  {Kouveliotou}, {Burud}, {Rhoads}, {Masetti}, {Palazzi}, {Pian}, {Pedersen},
  {Kaper}, {Gilmore}, {Kilmartin}, {Buckle}, {Seigar}, {Hartmann}, {Lindsay},
  \& {van den Heuvel}}]{vel+04}
{Vreeswijk}, P.~M., {et~al.} 2004, \aap, 419, 927

\bibitem[Vogt et al.(1994)]{hires} Vogt, S.~S., et al.\ 1994,
\procspie, 2198, 362

\bibitem[Wolfe et al.(1986)]{wolfe86} Wolfe, A.~M., Turnshek, D.~A.,
Smith, H.~E., \& Cohen, R.~D.\ 1986, \apjs, 61, 249

\end{thebibliography}
\end{document}